# Understanding BitTorrent Through Real Measurements

Wojciech Mazurczyk, Paweł Kopiczko
Warsaw University of Technology, Institute of Telecommunications
Warsaw, Poland, 00-665, Nowowiejska 15/19
Email: wmazurczyk@tele.pw.edu.pl, P.Kopiczko@stud.elka.pw.edu.pl

**Abstract.** In this paper the results of the BitTorrent measurement study are presented. Two sources of BitTorrent data were utilized: meta-data files that describe the content of resources shared by BitTorrent users and the logs of one of the currently most popular BitTorrent clients – μTorrent. μTorrent is founded upon a rather newly released UDP-based μTP protocol that is claimed to be more efficient than TCP-based clients. Experimental data have been collected for fifteen days from the popular torrent-discovery site *thepiratebay.org* (more than 30,000 torrents were captured and analyzed). During this period the activity and logs of an unmodified version of μTorrent client downloading sessions have been also captured. The obtained experimental results are swarm-oriented (not tracker-oriented as has been previously researched), which has allowed us to look at BitTorrent and its users from an exchanged resources perspective. Moreover, comparative analysis of the clients' connections with and without μTP protocol is carried out to verify to what extent μTP improves BitTorrent transmissions. To the authors' best knowledge, none of the previous studies have addressed these issues.

## 1. Introduction

BitTorrent [5], [3], a file transfer system originally released in July 2001, is currently the most popular P2P (Peer-to-Peer) networking system worldwide [8] that allows users to share particular resources in the form of files. Studies show that the number of users exceeded 100 million in 2011, and that BitTorrent traffic accounts for about 94% of all P2P traffic, and is responsible for about 22% of North American fixed access daily traffic [8], which is continuously increasing each year. This increase of the amount of BitTorrent traffic indicates that understanding the characteristics of BitTorrent can also improve the understanding of the overall Internet behavior.

The success of BitTorrent primarily comes from two factors: its efficiency and openness. BitTorrent is significantly more efficient than classical client/server-based architectures. It allows peers sharing the same resource to form a P2P network, and then it focuses on fast and efficient replication to distribute the resource. It is also worth noting that because in BitTorrent a resource is divided into many fragments, a single peer is able to download many fragments simultaneously and it does not need the whole resource to share it with other peers. Additionally, BitTorrent software is free to download and many clients' versions are open source. This leads to an easy deployment of new applications and technologies therefore stimulating further improvements.

The popularity and dynamic evolution of BitTorrent has led to an increased interest in its traffic measurements and analysis during recent years [4], [6], [7], [10], [11], [16], [17], [20] or [25]. The experimental results from these publications provide a good overview of BitTorrent users' behavior and resource "life cycle" in the network. However, they are lacking in several aspects: these studies were generally conducted some time ago, and the BitTorrent is constantly and dynamically evolving. For example, the studies do not account for usage of the most current and popular BitTorrent client – μTorrent [26] and its underlying transport protocol, – μTP (Micro Transport Protocol) [15], [26]. What is also worth noting is that the studies usually do not provide any information about the capabilities of the shared resources and how the BitTorrent network handles them.

As a result of the limitations in these previous studies, we elected to conduct and analyze new BitTorrent measurements. We focused on two sources from which BitTorrent information was acquired:

- Meta-data files (called torrent files) that describe the content of resources shared by BitTorrent users and are necessary to initialize connection to the network.
- μTorrent logs that are collected during resource downloading sessions. μTorrent is now the most popular BitTorrent client.

In order to carry out the study the data was collected for 15 days from one of the most popular torrent-discovery sites – *thepiratebay.org*. The activity and logs of forty-five μTorrent application downloading sessions have been captured. Data collected from both these sources was analyzed, and the obtained results are discussed in this paper.

Measurements used in this paper are based on the idea that meta-data files and client logs are critical sources of the BitTorrent data. The basic methodological concept is similar to the ones used in existing studies (see Section 3); however, the obtained results provide new insights. They enable us to explore BitTorrent from the users' perspectives, as well as through the resources that are actually shared by these users. Moreover, this methodology showcases how BitTorrent handles users' resources by introducing the concept of fragmentation. It also provides an important means through which to analyze the data that is characteristic for a single swarm (not for the whole tracker). By using an unmodified version of the popular client, µTorrent, we eliminated potentially negative influences that modified clients can have on the experimental results.

Therefore, through this methodology it is possible to observe what clients in a swarm are the most popular, and how many connections an unmodified µTorrent client establishes with default configurations. Additionally, chosen client software utilizes µTP protocol, which has not yet been subjected to the measurements incorporated in previous studies.

The key questions that we wish to answer in this paper are:

- What are the characteristics about the users' resources and how do BitTorrent clients handle them (in terms of the number of fragments in resources and resource fragment size)?
- Are µTP-based clients really more popular and capable of outperforming older TCP-based clients?

To authors' best knowledge, none of the previous studies have addressed these issues. Additionally, updated view on currently the most popular BitTorrent clients available and most popular resources is presented.

Measuring and monitoring network traffic is required to manage today's complex Internet backbones. Such measurement information is essential for short-term monitoring (e.g., detecting denial-of-service attacks), longer term traffic engineering (e.g., rerouting traffic and upgrading selected links), and accounting (e.g., to support usage based pricing). That is why it is especially important to perform measurements on BitTorrent because its traffic accounts for about 94% of all P2P traffic, and is responsible for about 22% of North American fixed access daily traffic and it has potentially large impact on the Internet performance.

The results of this paper should benefit diverse communities including P2P researchers, ISP researchers and copyright holders etc. Generally, there are four main reasons for the usefulness of network traffic measurements [6]: network troubleshooting, protocol debugging, workload characterization and performance evaluation. Results provided in this paper contribute in the last two. The development of Internet-equivalent workloads would provide the ability to engineer better systems. It would allow for test system modifications to be done in a controlled environment without disturbing real systems.

Depending upon the domain, network traffic measurements can serve different purposes. For example, an ISP (Internet Service Provider) can benefit from measuring the amount of outgoing traffic to estimate pricing and the services provided. On the other hand, manufacturers of network hardware (e.g. routers and switches) utilize real-world measurements to test the behavior of the hardware under realistic conditions without deploying them.

Thus, the main goals of this paper are to understand the characteristics of the BitTorrent system which can help in understanding users' needs, dimensioning of the ISP network or modeling P2P systems (e.g. for a P2P simulation environments).

The rest of the paper is structured as follows: Section 2 describes BitTorrent basics. Section 3 focuses on presenting existing work on BitTorrent measurements. Section 4 describes measurement methodology. In Section 5, experimental results are presented and analyzed. Finally, Section 6 concludes our work.

## 2. BitTorrent Basics

BitTorrent is a peer-to-peer file sharing system that allows its users to distribute large amounts of data (especially large files) over IP networks. What distinguishes BitTorrent from other similar file-transfer applications is that instead of downloading a resource (one or more files) from a single source (e.g. a central server), users download fragmented files from other users at the same time. In result, the file transfer time is considerably decreased because the group of users that shares the same resource (or part of it) may consist of several to thousands of hosts. Such group of users interested in the same resource (known as "peers") combines together with a central component (known as a "tracker") in BitTorrent. This combination of peers and trackers is called a "swarm" (Fig. 1). Trackers are responsible for controlling the resource transfer between the peers. Peers that hold onto a particular resource or part of a resource are required to share the resource and to perform the transfer.

We can distinguish two types of BitTorrent peers based on of the stage at which they are involved in downloading or sharing a given resource. These types are:

- Seeds – peers that possess the complete resource and are only sharing it.
- Leechers – peers that do not possess complete resource but they are interested in doing so. They also share the fragments they have already downloaded. When a leecher obtains all remaining fragments of the resource it automatically becomes a seed.

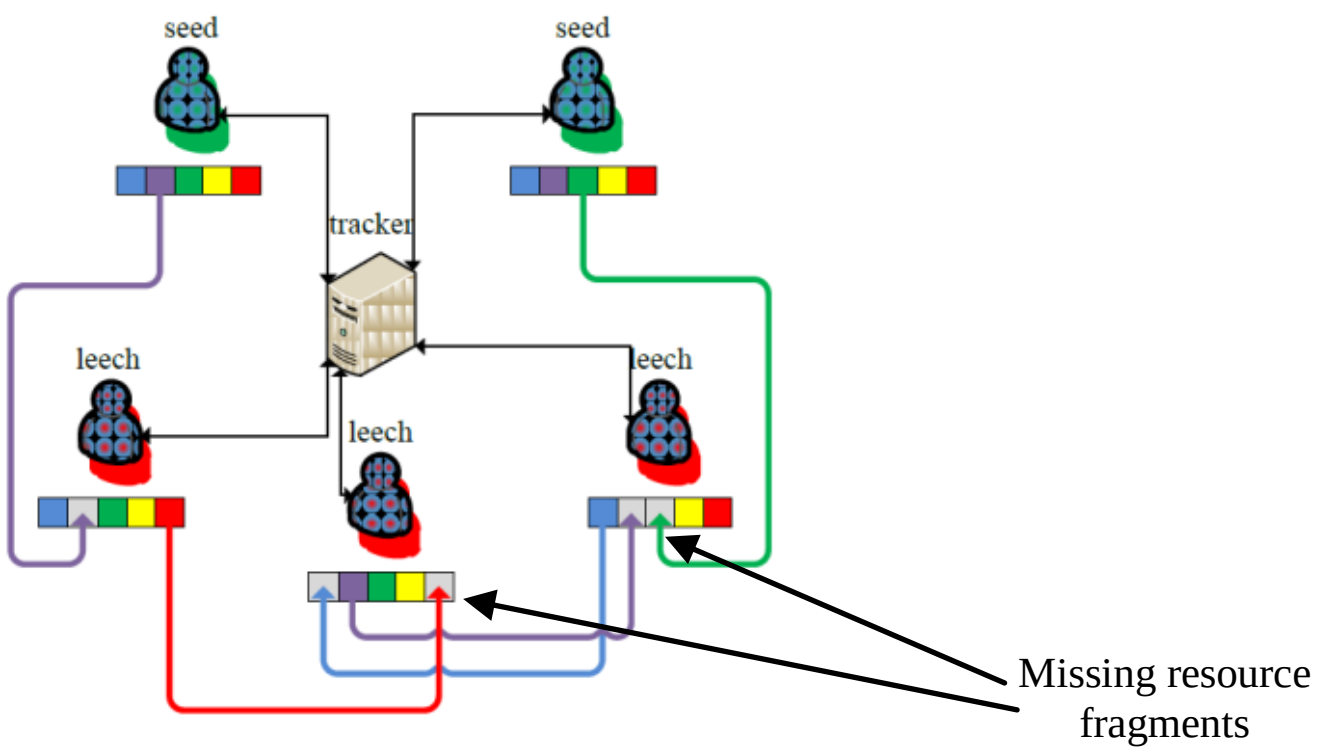

**Fig. 1** BitTorrent swarm resource sharing scenario (squares denote resource´s fragments)

A swarm can be controlled by at least one tracker. Tracker(s) are not aware of the resource content; they only know which peers currently have the complete file, or particular parts of a file or resource. Therefore, in order to initiate resource downloading it is not enough to know a tracker's URL. If a peer wants to join the swarm it must be in possession of special meta-data file with a *.torrent* extension that allows proper initiation of the data transfer (called torrent or meta-data file). BitTorrent specification does not imply the way the torrent files should be distributed among users. Usually, meta-data files are downloaded from one of the indexing servers that store them, and they enable a user to search for torrents based on the resource's description. The content of the meta-data file is bencoded [6]. Each resource that is shared in a BitTorrent network is fragmented into pieces; thus it is possible to break and reinitiate resource transfers without losing data acquired during previous sessions. In the event of a transmission error, it is also possible to verify parts' integrity (based on the parts' hashes that are present in meta-data file) and retransmit only the missing parts of the resource.

In BitTorrent specification ([5], [3]) two main protocols are described that regulate data transfer: *peer-tracker* and *peer-peer*.

The connection between peer and tracker can be established with the use of a HTTP (Hypertext Transfer Protocol), or through UDP-based requests. It must be emphasized that currently the role of the tracker diminished. The tracker is used mostly to initiate the connection with a swarm. After the connection is established, popular BitTorrent extensions like PEX (Peer Exchange) [3] or DHT (Distributed Hash Table) [13] are used. These extensions enable the communication between peers without using a tracker as part of the swarm.

In a BitTorrent specification, peer-peer data exchange should be conducted using an application layer – a proprietary TCP-based protocol. It is a stateful protocol that is used to establish connections similar to the TCP handshake mechanism. In fact handshake messages are the first messages exchanged between BitTorrent clients. If the exchange is successful then transmission of user content can take place, if not then such connection is considered as incomplete.

However, it is also possible that instead of using TCP in a transport layer, UDP-based µTP (µTorrent Transport Protocol) is used (Fig 2) that was introduced in 2009. It is not part of the original BitTorrent specification, but it was created by BitTorrent Inc., and as the results of this study illustrate, it is currently the most popular choice.

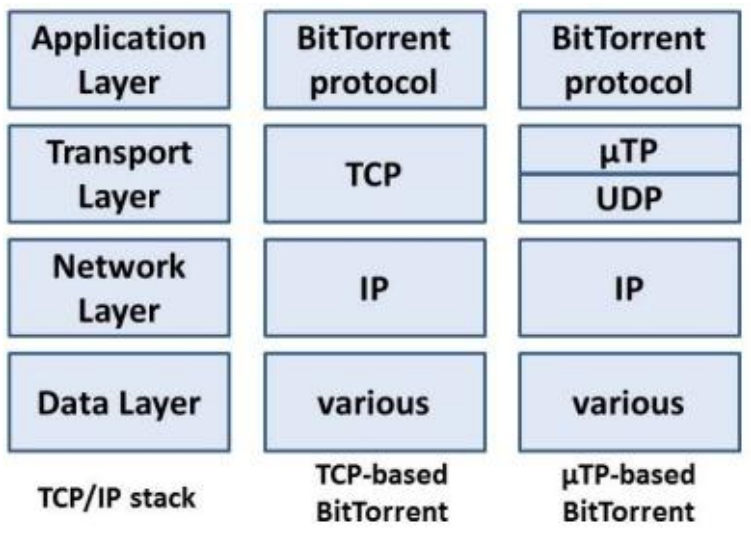

**Fig. 2** BitTorrent stacks for peer-peer connections

The main aim of the µTP protocol is to efficiently manage usage of the available bandwidth during file transfers, while limiting the impact of file transfers on the ongoing transmissions (especially not BitTorrent related ones). The µTP protocol is capable of automatically reducing the rate at which BitTorrent packets are transmitted between peers in case there is interference with other applications running on the same host. This protocol uses a congestion control algorithm, which is a modified version of LEDBAT (Low Extra Delay Background Transport) [19], based on one-way delay measurements. µTP protocol was implemented in the very popular BitTorrent client, µTorrent [26] beginning with version 2.0, and it is currently used by default. This protocol is also used in other BitTorrent clients, such as in BitTorrent [2], Vuze [22], Transmission [21], or KTorrent [12].

## 3. Related Work

Many P2P popular services have been targeted by scientific community and many different measurement studies were performed e.g. for Skype [1], [14] or other peer-to-peer systems [9], [24]. In the last few years also BitTorrent file transfer system has been the focus of numerous studies that were conducted by measuring real-life traffic [6], [7], [10], [11], [16], [17], [20] or [25]. These studies have provided interesting insights into BitTorrent networks and their characteristics. In general, existing studies can be divided into four groups based on what source is used to acquire information about BitTorrent:

- *Tracker as a source* [7], [10], [11], [16] – this is the most popular method that is utilized to obtain BitTorrent data, and it is based on capturing and analyzing tracker information. Two approaches are common: the first utilizes trackers' Scrape mechanism, which allows to ask the tracker about the status of the torrent file that is currently under its control; and the second approach is focused on analyzing the trackers' logs. These measurement methods allow researchers to identify the number of leechers and seeds in a given swarm; the number of global peers; the average client session download and upload rates; swarm lifetime length; and popularity of the torrent.
- *Meta-data files as a source* [11], [20], [25] – this type of study capitalizes on the idea that meta-data files are downloaded from indexing servers and are subsequently subject to analysis. This measurement method can be used to obtain information about: the popularity and categories of torrents; existing and active trackers; clients that create meta-data files [25]; and the time dependencies between published meta-data files.
- *Modified BitTorrent client as a source* [6], [20] – this measurement method provides researchers with the opportunity to analyze the BitTorrent swarms. This method is often related to analysis of meta-data files because the modified version of the client mostly connects to swarms by using meta-data files that were downloaded beforehand. Acquired information can give insights about: the average downloading and uploading data rate, and the number of leechers and seeds for the swarms in its range (i.e. those that are shared by a tracker, or those that can be found using BitTorrent protocol extensions like PEX or DHT). Modified clients can be also used to inspect BitTorrent users' anomalies, such as by identifying those users that do not allow any connections, or by analyzing many clients that hold the same IP addresses [20].
- *BitTorrent traffic captured by ISP as a source* [6], [7], [18] – this is the analysis of the traffic captured by an ISP. Such measurements are usually not limited solely to BitTorrent traffic, but they allow for inspection of the torrent packets' size and frequency, torrent packets' extensions, the amount of shares a packet has undergone across the Internet, and any changes or categorizations to the files that may occur among the most popular resources.

The results of the existing studies are usually difficult to compare because they were carried out some time ago and the BitTorrent trackers' and clients' software have dynamically evolved or changed throughout the years. Some currently popular clients did not exist when some studies were carried out (e.g. µTorrent). Moreover, as mentioned in Section 2, the role of the tracker is now much more limited, and it is not as commonly used in BitTorrent file sharing as it was few years ago. In addition, no studies have yet to present the influence of the µTP protocol on BitTorrent transmissions. That is why it is important that an updated picture of the BitTorrent network be presented, by performing new measurements that will capture some of the most important changes that have occurred in this resource-sharing system.

## 4. Measurement Methodology

In this study the measurement of BitTorrent is conducted based on two BitTorrent data sources: meta-data files and µTorrent client. Meta-data information is acquired from the *thepiratebay.org* indexing server, and the data from the µTorrent client is obtained from the client's logs.

The measurements of both these sources consist of the following three steps:

1. Experimental data acquisition;
2. Data analysis and data insertion to the relational database; and
3. Executing BitTorrent SQL query-based reports.

BitTorrent measurements based on meta-data files works as follows: *thepiratebay.org* server is also a tracker in which URL addresses are: *http://tracker.thepiratebay.org/announce* and *udp://tracker.thepiratebay.org:80/announce*. We benefit from this situation because this allows for the acquisition of information from web pages of this indexing server. The acquired data pertains to the number of seeds, the number of leechers in a swarm for each torrent, the URL address of the meta-data file, and other additional data like the torrent file creation date and the name of the user that added it. Specialized applications send HTTP requests to the *thepiratebay.org* indexing server. The server then replies with the list of torrents that are subsequently sorted by the number of seeds. These lists are further divided into three categories: videos, games, and software. After receiving this information, the data is preprocessed and the following information about each torrent is inserted into database: category of the resource, number of seeds and leechers estimated by tracker, the meta-data file URL address, and any additional information about the torrent itself.

In the next step, the meta-data file is extracted and analysis of all new meta-data files is carried out. At this stage, the following new information is then added to the database: the torrent's name, resource size, tracker URL, URL trackers list from announce-list field, the number of fragments that the resource was divided into, the size of the fragments, the list of files in the resource, the torrent creation date, and any users' comments.

BitTorrent measurements based on the µTorrent client's logs are performed by extracting information about BitTorrent client connections. In our study, the unmodified version of µTorrent 3.0 was used (which is different when compared with existing approaches – see Section 3), and only the most popular resources are downloaded. The most popular resources are characterized as having considerably more than 100 seeds. The client's logs are then acquired during the resource downloading sessions.

For each connection the following data is stored in the database: initiation and ending date of the connection, its direction (incoming or outgoing), IP address and port number of the remote host, the BitTorrent client's name, whether µTP protocol was used, and if handshake mechanism was performed.

The client's logs may not include information about the connection ending and whether the handshake was performed. Such situations can occur when the limit to the number of connections in a BitTorrent client is exceeded; when logging was initiated when the downloading session had already been active; or when the logging was cancelled during active connections. It must be noted that if there was no handshake performed then no information can be extracted about the BitTorrent client used for the connection. Additionally, if the analyzed connection's direction is incoming, then it is not possible to find out whether the µTP protocol was meant to be used for a particular connection.

The presented measurements in this methodology section are based on developed, specialized applications – the utilized experimental environment is presented in Fig. 3.

Applications that were designed and developed to search the indexing server periodically sent requests and then receive and analyze HTML pages to extract the meta-data files (Fig. 3, 1). Applications were initiated in a single session for twelve hours, and twelve pages of torrent lists for each analyzed torrent category were downloaded. Requested data was sorted by the number of seeds. On each page there were thirty meta-data files with corresponding URL addresses presented together with other information pertaining to the torrents. Therefore a total of 1080 torrents were captured in each session. The data was acquired for fifteen days in July 2011. Therefore, more than 30,000 torrents were analyzed in this time period. After acquiring data from the indexing server, parsers were applied on HTML pages and meta-data (Fig. 3, 2), and then the desired information was saved in the database (Fig. 3, 3).

Logs from the unmodified version of the µTorrent client (Fig. 3, 4) were obtained by initiating three downloading sessions once a day, each lasting for fifteen minutes. Downloading sessions focused primarily on downloading the most popular resources (defined as torrents containing seeds significantly greater than 100). The data acquisition was carried out for a total of fifteen days. Therefore, logs from forty-five downloading sessions were captured and analyzed. Session-specific information, meta-data files, and application logs were stored during this time period. This data was further analyzed by a µTorrent log parser (Fig. 3, 5), and the extracted information was stored in the database.

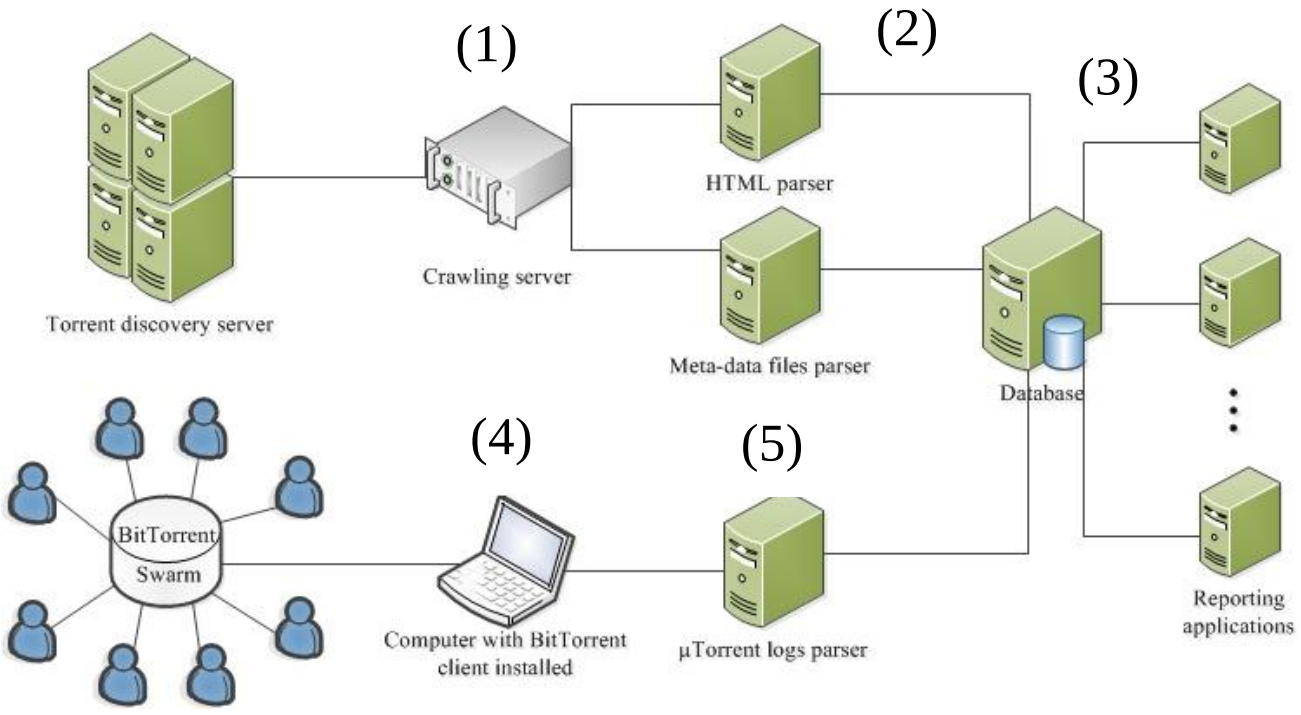

**Fig. 3** Experimental environment

## 5. Measurement Results

### 5.1 Analysis of meta-data files

Using the proposed methodology for the meta-data files described in Section 4, more than 30,000 torrent files were captured and analyzed. All files were obtained from one of the most popular BitTorrent indexing server, *thepiratebay.org*. Meta-data files were chosen based on their popularity (i.e. based on tracker's estimated number of seeds). Using the captured data we inspected the following five BitTorrent swarm characteristics:

- the number of seeds,
- the resource type and size,
- the resource fragments size,
- the number of files in the resource, and

- the number of resource fragments.

Capturing of this data allows to answer the question on what are the characteristics about the users' resources and how do BitTorrent clients handle them (in terms of the number of fragments in resources and resource fragment size).

Fig. 4 (left) presents the relationship between the number of torrents and the number of seeds that share them (ranging from 0 to 10,000). The maximum number of peers that shared the complete resource (seeds) was 40,058; however, for the sake of clarity, the diagram was limited to 10,000 peers as there were only a few single torrents with seeds significantly higher than that number.

Most often, resources were shared by approximately 150 to 400 seeds, with an estimated 50% of the analyzed torrents falling under this category. The average number of seeds for a given resource was about 1160. In a range from 0 to 1000, most torrents were shared by approximately 180 seeds (Fig. 4, right).

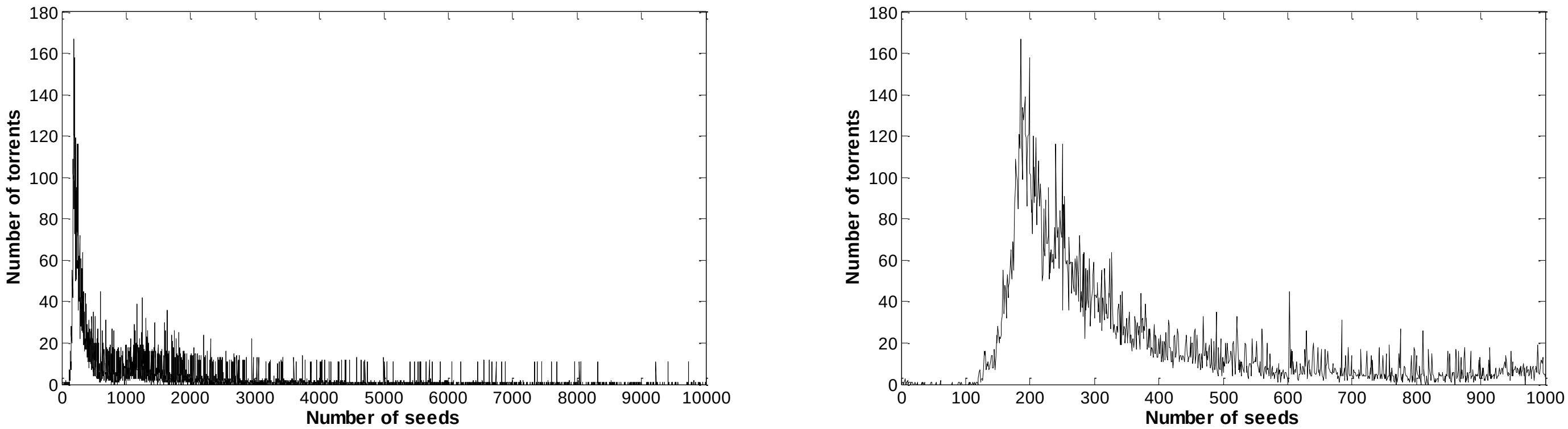

**Fig. 4** The number of torrents for seeds that share them in range 0 - 10 000 (left) and 0 - 1000 (right)

Another BitTorrent swarm characteristic that we wanted to inspect was torrent size (Fig. 5, left). About 98% of all analyzed meta-data files fit in the range between 0 GB and 10 GB. The largest resources (approximately 28 of them) were 28.5 GB in size or greater. These resources were typically software. The resources exceeding 10 GB in size are rather rare; however, resources that were approximately 8 GB were frequently encountered. It should be noted that the average size of resource was about 2 GB.

For resources less than 1 GB (Fig. 5, right), there were few interesting spikes in number of torrents between about 1 MB to 5 MB, 175 MB, 350 MB, 550 MB and 700 MB. The largest number of torrents (about 1700) occurred in the range of 349 MB to 350 MB, and these were mostly resources that represented episodes from popular TV series. Other spikes occurred at 175 MB, 550 MB and 700 MB, which were also caused by downloading videos. Spikes noted between approximately 1 MB to 5 MB were mostly related to the torrents related to small software (approximately 1300 in total), and single music files (about 300 in total).

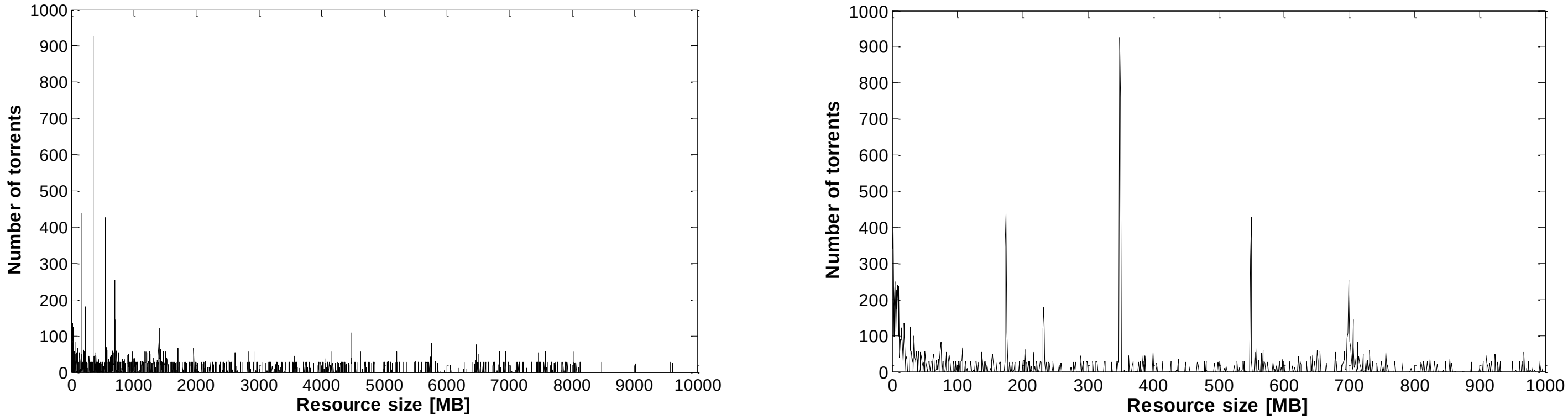

**Fig. 5** The resource size in range 0 MB to 10000 MB (left), and 0 MB to 1000 MB (right)

Fig. 6 (left) illustrates a number of files in each resource. Results presented in the figure were limited to 100 files, because about 96% of analyzed resources fall into this range. However, there were single meta-data files recorded that were related to resources consisting of approximately 4000 files (software). About 10,500 resources consisted of a single file (which totaled approximately 33% of all inspected torrents), and about 27,500 of resources consisted of fewer than 20 files. Resources with a higher number of files (i.e. greater than 20 files) were significantly less popular (Fig. 6, right).

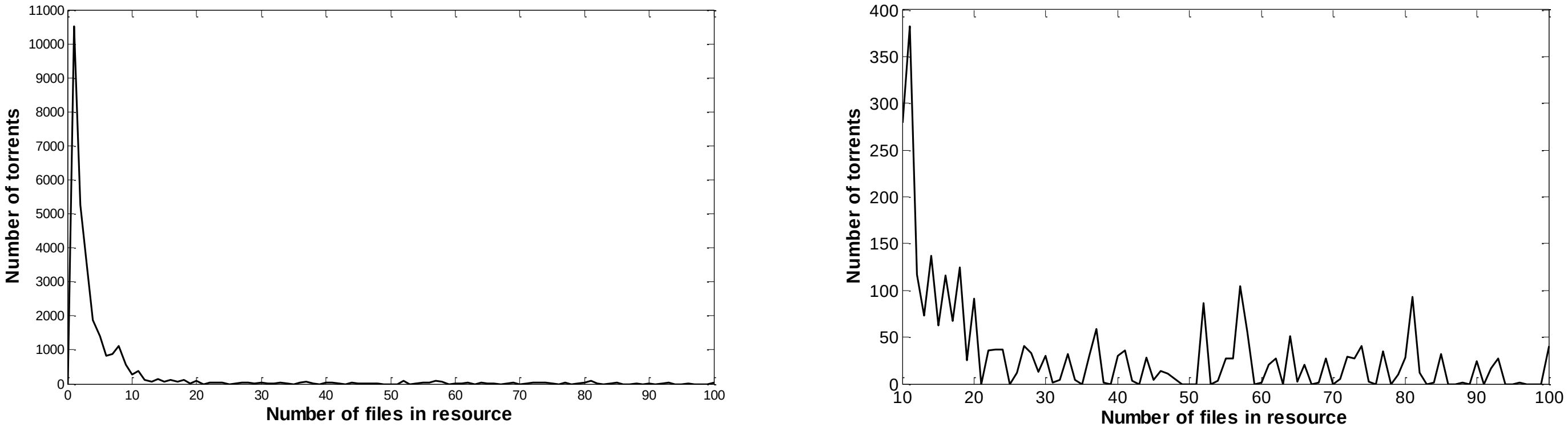

**Fig. 6** The number of files in resource in range 0 - 100 (left) and in range 10 - 100 (right)

Fig. 7 (1) illustrates resource fragment size characteristics. In all inspected meta-data files, resources were divided into fragment sizes of at least 16 KB, and not higher than 8092 KB. The most frequent fragment sizes were those that represented subsequent power of 2 (beginning from 16).

Fig. 7 (2) presents the number of fragments that resources were divided into within meta-data files. About 95% of all resources fell in a range from approximately 0 to 5000 fragments. However, single resources were divided into thousands of fragments, and the maximum value was above 51,000. The majority of resources were divided into 500 to 1000 fragments. We also observed a few spikes occurring at 1 to 2 fragments (totaling approximately 500 resources), 350 to 351 fragments (about 1100 resources), 699 to 701 fragments (about 2000 resources), and 1100 to 1101 fragments (about 500 resources).

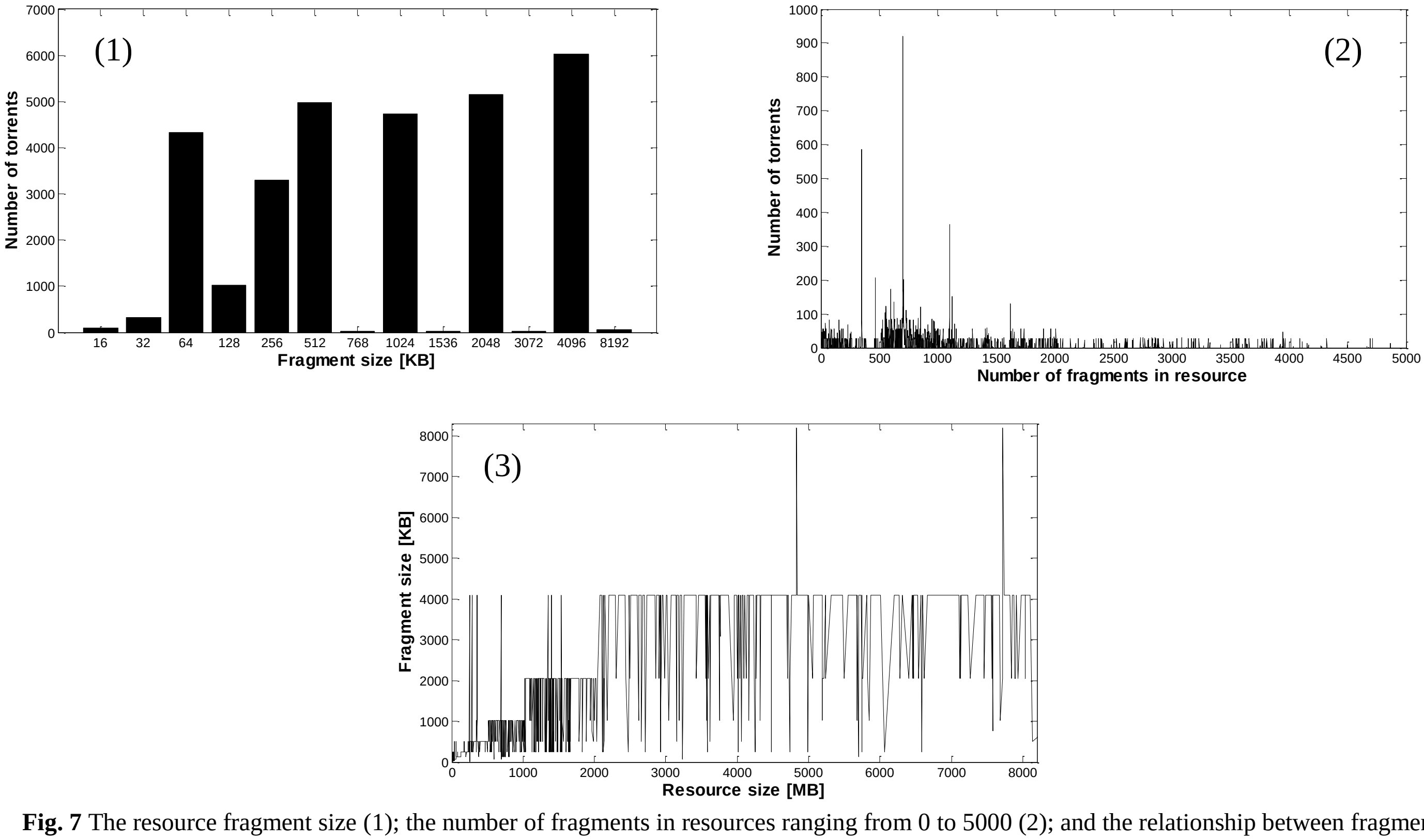

**Fig. 7** The resource fragment size (1); the number of fragments in resources ranging from 0 to 5000 (2); and the relationship between fragment and resource size (3)

Fig. 7 (3) describes relationship between resource and fragment size. Based on the obtained results it can be concluded that there is a link between resource size and size of each fragment comprising the resource. A resource size of up to 1 GB typically contains fragments measuring between 512 KB and 1024 KB. When the size of the resource is between 1 GB and 2 GB, then each fragment size is likely to be 2048 KB. For resources larger than 2 GB, each fragment is likely to be 4096 KB. Therefore, the larger the resource's size the larger the chosen fragment's size.

To summarize, from the experimental results noted above, one can conclude that resources are most likely shared by approximately 180 seeds. However, it should be noted that there are cases in which the number of seeds has exceeded several thousands. The largest number of seeds that share the same resource is 40,058, and the average number of seeds sharing a resource is 1160. Resources usually consisted of 1 to 12 files, and their sizes were often less than or equal to 10 GB, with characteristic spikes observed at about 1 MB, 175 MB, 350 MB, 550 MB, and 700 MB. Resources were typically divided into fragments sized 64 KB, 512 KB, 1024 KB, 2048 KB, or 4096 KB. The number of these fragments was typically equal to 1, 2, 350, 700, or 1100.

### 5.2 Analysis of μTorrent client logs

Using the methodology for analyzing the μTorrent client that was described in Section 4, the logs from forty-five downloading sessions were captured for popular torrents, and fifteen minutes of each session were analyzed. Information pertaining to the time of each initial connection and termination were saved.

This data was collected to find out whether μTP-based clients are really more popular and capable of outperforming older TCP-based clients. Additionally, data pertaining to each connection (incoming or outgoing) was also captured to answer the following questions:

- Was the connection established with a use of μTP protocol?
- Was there handshake performed?
- What BitTorrent client was utilized? (This analysis was carried out only for connections that successfully performed a handshake)

Table 1 presents the level of popularity for the BitTorrent client's users, and is characterized by the number of connections that were established. It can be seen that the most popular clients are μTorrent and BitTorrent. Together they dominate all clients, as they comprise a total of 93% of file sharing. Interestingly, both of these clients are products of BitTorrent Inc. Less popular clients were often used in a Linux environment. These clients included Transmission and Azureus (currently named Vuze), and they comprised a total of 4.5% of all connections. Other clients were used rather rarely in comparison.

**Table 1** BitTorrent client popularity

| Client | μTorrent | BitTorrent | Transmission | Azureus (Vuze) | Bit Comet | Lib torrent | Ares | Unknown | Deluge Torrent | KTorrent | Xunlei | Bit Lord |
|---|---|---|---|---|---|---|---|---|---|---|---|---|
| [%] | 72.05 | 20.61 | 2.44 | 2.08 | 1.8 | 0.28 | 0.13 | 0.13 | 0.12 | 0.09 | 0.07 | 0.02 |

Fig. 8 (left) illustrates the average number of all captured connections - those in which the handshake was performed, and those that utilized μTP protocol. In each case, the number of client connections increased for approximately three to four minutes in the downloading session, after which point it began to stabilize at a certain level. If we considered all connection, this number stabilized at about 80 connections.

It was also observed that during the first minute of a downloading session, the number of all connections was significantly greater than the number of connections in which the handshake was performed. The difference (about thirty connections) was visibly greater than in the rest of the downloading sessions (when the difference is about ten connections). This discrepancy is caused by many connection attempts with other BitTorrent clients that are refused. In other words, the μTorrent client is trying to find the largest number of available sources for the desired resource prior to downloading the corresponding files.

The number of connections with a handshake stabilizes due to the μTorrent client's default number of allowed connections (i.e. 100 connections for a single downloading session), or due to a user reaching his or her Internet connection bandwidth limit (in this experimental environment, the limit was 6 Mb/s).

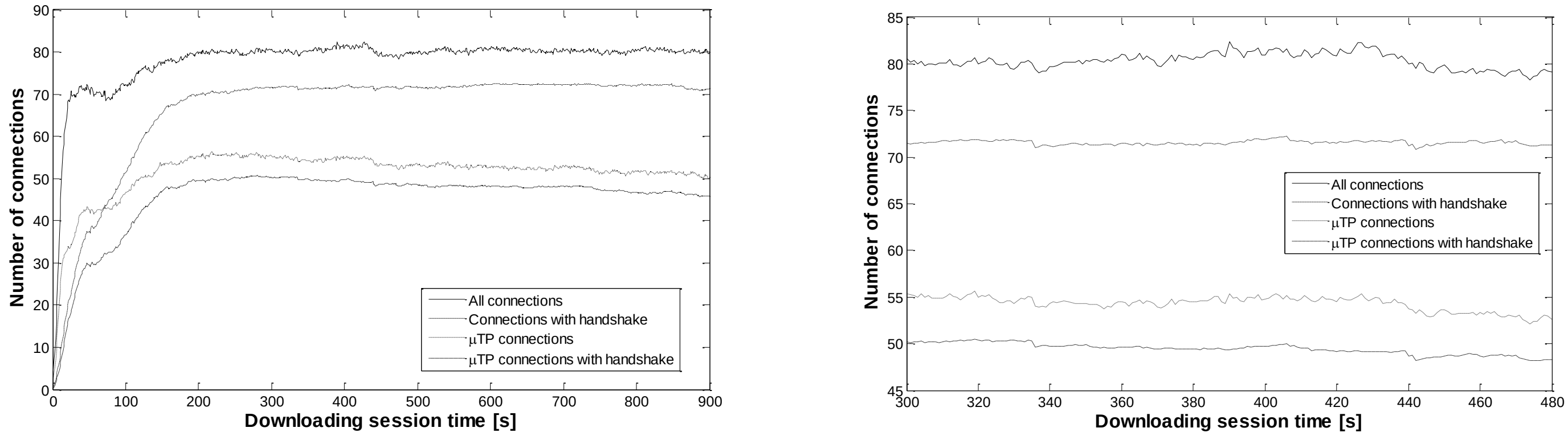

**Fig. 8** Number of connections across all analyzed downloading sessions using μTorrent (left), and between five and eight minutes (right)

It is also worth noting that the number of all captured connections in which handshake was performed was about 22,000, and the number of all registered connections reached above 264,000. Therefore, it can be concluded that from the fifth minute of the downloading session, the difference in the number of connections with and without a handshake is caused by a large number of a short-lived connections. This phenomenon was observed as an irregularity of the total number of connections curve shown in Fig. 8 (right). This is further confirmed in Fig. 9.

Next, let us focus on inspecting the connections established using µTP protocol. Generally, it can be observed that a characteristic of all µTP connections and µTP connections with a handshake is that their data curves behave similarly. µTP protocol was used to established connections in 66% of the cases, which illustrates the popularity of this protocol among users and clients alike. It is also worth noting that, if we were able to analyze the number of refused connections, the popularity of the µTP protocol might be even greater. Furthermore, the irregularities in the number of µTP connections were lower when compared with the total number of connections (Fig. 8, right). Finally, the difference between all µTP connections and µTP connections with a handshake is lower (of about 5 connections), which proves that the µTP is superior to the case when TCP is used in a transport layer (where the difference is about 10).

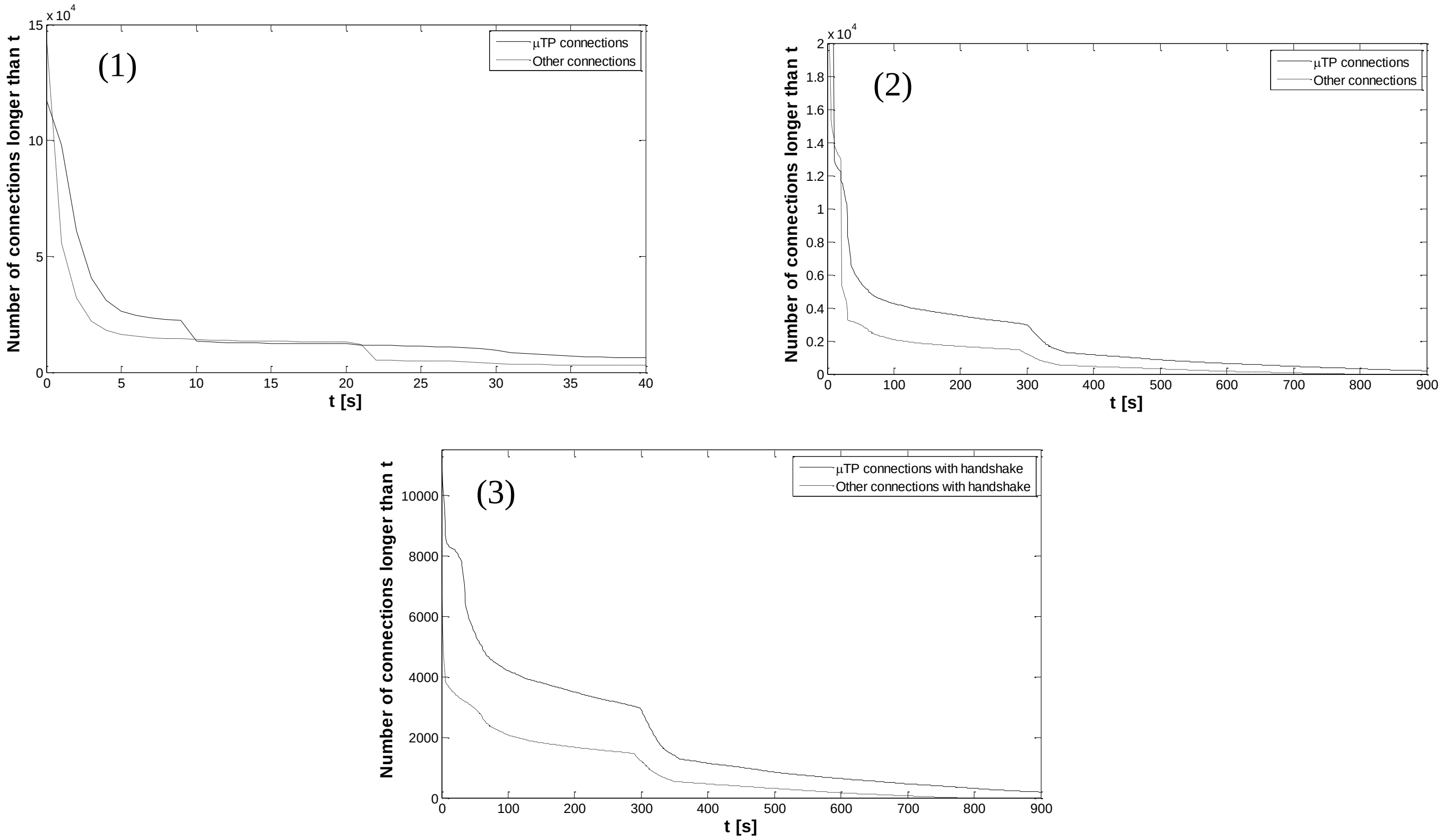

**Fig. 9** The number of connections during downloading sessions using µTorrent lasting longer than t for all connections from 1 s to 40 s (1), and all connections ranging from 0 s to 900 s (2), and connections with a handshake ranging from 0 s to 900 s (3)

Fig. 9 (1) further confirms the above conclusions. When we take into account all the connections (with and without a handshake), we can observe that the number of short-lived TCP-based connections, that last less than 1 s, is larger (totaling about 88,000 connections) than for µTP-based ones (totaling about 19,400 connections). It should be noted that the number of µTP-based connections ranging from 1 s to 10 s, and from 10 s to 20 s also appear to be similar. In addition, there are more µTP-based connections that last longer than 20 s, and an even greater number of connections last for the remainder of the downloading session (Fig. 9 (2)). This is further highlighted in connections with a handshake, where the number of connections in a µTP-based protocol dominates the entire downloading session time. However, the difference is smaller for connections that last longer than five minutes.

To summarize, based on µTorrent client logs we can conclude that during the resource downloading session a large number of connections is established. However, in a majority of these sessions the connections are refused due to users exceeding the allowed number of simultaneous connections configured in the client. It was also observed that during the first three to four minutes of a downloading session, the number of established connections increased until it reached a relatively stable level of about 80 connections. Generally, the connections with handshakes dominated each time, despite the fact that their numbers were lower when compared to connections without handshakes. In general, connections with handshakes lasted longer than other connections.

Therefore, the most popular BitTorrent clients are currently µTorrent and BitTorrent. Because both of them utilize µTP protocol for connection establishment, µTP was used by a total of 66% of all connections with a handshake.

## 6. Conclusions and future work

The purpose of this paper is to help in the understanding of a real and currently the most popular P2P system – BitTorrent – by providing measurement data that can be useful, for example, in helping understanding users' needs, dimensioning the network or modeling P2P systems. Measuring and monitoring network traffic is required to manage today's complex Internet backbones. That is why it is especially important to perform measurements on BitTorrent because its traffic accounts for about 94% of all P2P traffic, and is responsible for about 22% of North American fixed access daily traffic and it has potentially large impact on the Internet performance.

The measurement study that was presented in this paper was focused on BitTorrent resources that are exchanged by users and the clients that are used to connect to P2P networks. An interesting finding was that there were two popular clients: µTorrent and BitTorrent, which are used in 93% of all downloading sessions. This finding demonstrates that the µTP protocol (which is used in both of these clients) is an important component of the BitTorrent network. As illustrated, clients that utilizing µTP behave better than other TCP-based clients. These clients resulted in a lower overall difference between all established connections, as well as those with successful handshakes. Moreover, the first few minutes of the downloading session were characterized by a marked increase in the number of initial connections, while connections began to level off to approximately 80 connections after three to four minutes.

The resources in BitTorrent were usually shared by about 180 seeds, and they consisted of between one to twelve files, with sizes of up to approximately 2 GB. To improve a resource's transmission, the files were typically divided into a number of fragments (usually between 350 and 1100 files), with file sizes ranging from 64 KB to 4096 KB.

Future work will include conducting further studies on µTP-based BitTorrent clients. In particular, we would like to find out if µTP-based clients are favoring other, less popular µTP-based clients, as was originally assumed with the advent of µTorrent version 2.0. Additionally, a more in-depth analysis of µTP performance for different BitTorrent clients is required in order to establish a strong comparison with TCP-based clients.

Moreover, based on the results in this paper the derivation of the most suitable model for BitTorrent traffic could be achieved. Traffic modelling is an important activity in the context of predicting future network behaviour. Accurate models for various networked applications are a decisive step towards QoS in the Internet. With the rise in popularity of P2P protocols, the importance of tractable and practically usable models is further increased. The derivation of the most suitable model for BitTorrent based on performed experimental results is considered also our future work.

## ACKNOWLEDGMENT

This research was partially supported by the Polish Ministry of Science and Higher Education and the Polish National Science Center under grants: 0349/IP2/2011/71 and 2011/01/D/ST7/05054.